\documentclass[aps,pra,twocolumn,amsmath,amssymb,nofootinbib,showpacs,superscriptaddress]{revtex4-1}
\usepackage[english]{babel}
\usepackage{latexsym}
\usepackage{graphics}
\usepackage{graphicx}
\usepackage{epsfig}
\usepackage{color}
\usepackage{bm}
\usepackage{amsmath}
\usepackage{amssymb}
\usepackage{amsthm}
\usepackage{dcolumn}
\usepackage{bm}
\usepackage{float}
\usepackage{hyperref}
\usepackage{color}
\usepackage{epstopdf}
\usepackage{cleveref}
\usepackage[svgnames]{xcolor}
\hypersetup{hidelinks,colorlinks=true,allcolors=DarkBlue}
\begin{document}

\title{Post-processing procedure for industrial quantum key distribution systems}
\author{E.O. Kiktenko}
\affiliation{Theoretical Department, DEPHAN, Moscow 143025, Russia}
\affiliation{Bauman Moscow State Technical University, Moscow 105005, Russia}
\author{A.S. Trushechkin}
\address{Steklov Mathematical Institute of Russian Academy of Sciences, Moscow 119991, Russia}
\affiliation{Theoretical Department, DEPHAN, Moscow 143025, Russia}
\author{Y.V. Kurochkin}
\address{Russian Quantum Center, Moscow 143025, Russia}
\author{A.K. Fedorov}
\affiliation{Theoretical Department, DEPHAN, Moscow 143025, Russia}
\address{Russian Quantum Center, Moscow 143025, Russia}

\date{\today}
\begin{abstract}
We present algorithmic solutions aimed on post-processing for industrial quantum key distribution systems with hardware sifting. 
The main steps of the procedure are error correction, parameter estimation, and privacy amplification. 
Authentication of a classical public communication channel is also considered.
\begin{description}
\item[PACS numbers]
03.67.Dd, 03.67.Ac, 42.50.Ex
\end{description}
\end{abstract}
\maketitle

\section{Introduction}

Significant attention to quantum key distribution \cite{GisinScarani} is related to the fact of breaking of public-key encryption algorithms using quantum computing.
Security of public-key exchange schemes can be justified on the basis of the complexity of several mathematical problems.
Nevertheless, the Shor's algorithm \cite{Shor} allows solving these problems in a polynomial time. 
Absence of efficient classical (non-quantum) algorithms breaking public-key cryptosystems still remains unproved. 

In view of possibility to establish a shared private key with unconditional security between two users (Alice and Bob) via quantum key distribution \cite{GisinScarani}, 
using of information-theoretically secure one-time-pad encryption technique becomes a practical tool.
Privacy of quantum keys is guaranteed by the laws of quantum physics \cite{Wootters}.
Quantum key distribution has been realized in experiments \cite{BB2,Shields,Weinfurter,Gisin2,Gisin3}. 
Devices for quantum key distribution are available on the market \cite{Market}. 

On the top of using of single quantum objects (photons) as information carriers, 
in quantum key distribution protocols such as the seminal BB84 protocol \cite{BB84} classical communications and post-processing are required \cite{BB84}.
A key requirement for quantum key distribution is that classical communications are not distorted.
In other words, an eavesdropper is able to obtain information from this channel, but cannot to change it.
To this end, one can use authentication schemes.
Due to the technological limitations of quantum key distribution, keys of Alice and Bob are initially different even in the absence of eavesdropping.
Excluding of this effect can be realized by using error correction procedures.
On the basis of compassion of keys after error corrections, one can estimate the quantum bit error rate (QBER). 
If QBER exceeds a critical value, the parties receive warning message about possible of eavesdropping.
In other cases, they use privacy amplification methods to exclude announced information on previous stages from keys. 

In this contribution, 
we report about the joint research project, which is aimed on a design of an industrial fiber based quantum key distribution system in Russia in collaboration between four teams.
The quantum key distribution engine is based on the decoy states BB84 protocol \cite{Decoy}. 
We present the developed post-processing procedure for sifted quantum keys ({\it i.e.}, keys after basis and intensity reconciliations), 
which consists of the following steps: error correction, parameter estimation, and privacy amplification.
Communications over public channel are authenticated.

In Sec.~\ref{sec:setup}, we briefly describe the hardware engine of our system.
In Sec.~\ref{sec:procedures}, we consider post-processing algorithms for error correction, parameter estimation, and privacy amplification.
We discuss authentication of public channel in Sec.~\ref{sec:authentication}.
The workflow of the post-processing procedure is presented in Sec.~\ref{sec:workflow}.
We summarize our results in Sec.~\ref{sec:conclusion}.

\section{Quantum key distribution setup}\label{sec:setup}

Our setup for quantum key distribution consist of two modules on the each side (Alice and Bob).
First, control units, that are connected via the optical fiber and perform all operations with photons.
Second, conjugation units, that are connected with the classical public channel, perform all post-processing, and export final keys to external applications.  

Operation of the control units is based on the ``plug and play''~\cite{Muller} principle of realization of the BB84 protocol with decoy states~\cite{Decoy}.
Control units work as follows.
Bob sends a sequence (``train'') of strong coherent pulses to Alice.
On her side these pulses are (i) reflected by a Faraday mirror, 
(ii) phase modulated with four possible values according to random basis and bit value, 
(iii) attenuated to the on three intensities (vacuum, signal, or decoy), 
and (iv) sent back.
The scheme passively compensates slowly varying polarization fluctuations in the fiber.
In the Bob's control unit, the returned attenuated pulses are phase modulated with two possible values that correspond to random choice of basis, 
and measured by pair of single photon detectors.

After an each train of pulses, Bob's conjugation unit obtains time indices of detected signals and corresponding bit and basis values.
These time indices and values of basis choices are sent to Alice's conjugation unit via classical channel.
Alice compares the measurement bases with preparation bases and sends Bob time indices of signal bits, 
which have been measured and prepared in the same basis. 
This procedure is know as sifting.
The values of signal qubits that were prepared and measured in the same basis give so-called sifted keys, that we denote as $K^{A}_{\rm sift}$ and $K^{B}_{\rm sift}$.
We note that Alice's conjugation unit also obtains rate of detection events for all three intensities that are used for revealing of photon-number splitting attack.  
However, in this work we restrict ourselves to post-processing without paying attention to analysis of decoy states.

We also note that random choices of bit values and preparation basis in Alice's control unit and random choices of measurement bases in 
Bob's one are performed using random number generators.
Random number generators are important and often forgotten ingredient of quantum key distribution systems \cite{Gisin3}.
Their development is a part of the project \cite{Chizhevsky}.
Random number generator is used several times during the protocol: 
(i) to generate random photon states, 
(ii) to generate random information specifying the verification hash function, 
and (iii) to generate random information specifying the hash function in verification (part of error correction) and the privacy amplification procedure.

\section{Algorithms for the post-processing procedure}\label{sec:procedures}

In this section, we consider processing procedures of two sifted quantum keys $K^{A}_{\rm sift}$ and $K^{B}_{\rm sift}$.

\subsection{Error correction}

The error correction algorithm is applied for making the sifted quantum keys $K^{A}_{\rm sift}$ and $K^{B}_{\rm sift}$ to be identical on the both sides. 
We employ the following error correction algorithm with two basic stages.
The first is to use the low-density parity-check (LDPC) syndrome coding/decoding \cite{LDPC} to correct discrepancies between keys.
The universal polynomial hashing \cite{Krovetz} to verify an identity between keys after previous step is used as the second stage. 

Alice and Bob share a pool of LDPC parity-matrices with code-rates $R$ from 0.9 up to 0.5 (with step being equal to 0.05) and the frame size (length of processed strings) being equal to $n=4096$.
These matrices are constructed with progressive edge-growth algorithm~\cite{Reconciliation2} using polynomial (degree distributions) from Ref.~\cite{Reconciliation3}.
For each coding and decoding process parties employ code with the minimal rate $R$, which satisfies the following condition:
\begin{equation}
	\frac{1-R}{h_\mathrm{b}(\mathrm{QBER}_\mathrm{est})}\leq f_\mathrm{crit}
\end{equation}
where $h_\mathrm{b}$ is the standard binary entropy function, 
$\mathrm{QBER}_\mathrm{est}$ is the estimated level of QBER (see below), 
and $f_\mathrm{crit}=1.22$ is the critical efficiency parameter in our setup, 
that is the tolerable ratio between level of disclosed information about sifted key and theoretical limit for successive error correction,
which is predicted by the classical information theory.

To decrease a frame error rate (probability of unsuccessive decoding) in the constraint that the resulting efficiency is not greater than $f_\mathrm{crit}$,
Alice and Bob use the shortening technique~\cite{Reconciliation}. 
The number of shortened bits $n_s$ is obtained from the following expression
\begin{equation}
	n_s=\lfloor{n-\frac{m}{f_{\rm crit}{h_\mathrm{b}(\mathrm{QBER}_\mathrm{est})}}}\rfloor,
\end{equation}
where $\lfloor x \rfloor$ stands for the maximal integer less than $x$.
Alice and Bob construct $\mathcal{N}=256$ strings of length $n$ possessing $n-n_s$ bits of their sifted keys $K^{A}_{\rm sift}$ and $K^{B}_{\rm sift}$ and $n_s$ shortened bits, 
whose positions and values come from synchronised pseudo-random generator.
Bob multiplies the chosen parity-check matrix on the constructed strings to obtain syndromes that are sent to Alice.
For the LDPC syndrome decoding, 
Alice applies iterative sum-product algorithm~\cite{Reconciliation}, which uses log-likelihood ratios for messages between symbol and parity-check nodes of the corresponding Tanner graph. 
To avoid costly calculations, we employ optimization techniques considered in Ref.~\cite{LDPC2}.
Then Alice removes shortened bits obtaining corrected key. 

If the algorithm of decoding does not converge for particular block in a specified number of iterations (we use about 60 iterations in our setup), then this block is considered as unverified.
However, rarely decoding process converges to a wrong result, {\it i.e.}, to a incorrect key but still with proper syndrome.
To avoid such situations, the second step of verification is used for blocks successively completed decoding.

In our setup, we employ comparison of hash-tags constructed with $\varepsilon$-universal polynomial hashing~\cite{Krovetz}.
In particular, 
we use modified $50$-bit variant of \textsf{PolyR} hash function that provide collision probability for a $\mathcal{N}$ blocks of $n-n_s$ bits on the level of $\varepsilon_\mathrm{ver}<2\times10^{-12}$.
If the hash tags on the both sides match, then the corresponding blocks are concerned to be identical on the both sides and are added to verified keys $K_{\rm ver}$.
We note that $K_{\rm ver}$ is a part of $K^{B}_{\rm sift}$ as all modifications of the key are performed on the Alice's side.

\subsection{Parameter estimation}

The purpose of the parameter estimation procedure is to determine QBER that is a probability of bit-flipping in a quantum channel. 
This problem could be resolved by comparison of input $K^{A}_{\rm sift}$ and output $K_{\rm ver}$ keys of the error correction, 
because as it was noted all changes of the key were performed exclusively by Alice. 
The ratio of corrected bits is averaged over a set of $\mathcal{N}$ the error correction blocks, and for unverified blocks a conservative maximal value $1/2$ is assumed.
If we denote all verified blocks in the set as $\mathcal{V}$ than the estimated QBER reads
\begin{equation}\label{QBER}
	\mathrm{QBER}_\mathrm{est}={\mathcal{N}}^{-1}\left(\sum\nolimits_{i\in\mathcal{{V}}}\mathrm{QBER}_i+\left|\overline{\mathcal{V}}\right|/2\right),
\end{equation} 
where $\mathrm{QBER}_i$ is the ratio of bit-flips in $i$th block and $|\overline{\mathcal{V}}|$ is number on unverified blocks.

\subsection{Privacy amplification}\label{sec:pa}

After these procedures, both sides have identical bit strings.
Nevertheless, Eve may have some amount of information about them. 
The privacy amplification procedure is used to reduce this potential information of an adversary to a negligible quantity. 
This is achieved by a contraction of the input bit string into a shorter string. 
The output shorter string is a final private key $K_{\rm sec}$ of length $l_{\rm sec}$. 
Our algorithm of the privacy amplification firstly checks (according to Ref. \cite{Tomamichel}) 
whether it is possible to distill private key with the given length and security parameter $\varepsilon_{\rm pa}$
(which is set to $\varepsilon_{\rm pa}=10^{-12}$). 
Namely, define the quantity
\begin{equation}
	\nu=\sqrt{\frac{2(l_{\rm ver}+k)(k+1)\ln\frac1{\varepsilon_{\rm pa}}}{l_{\rm ver}k^2}},
\end{equation}
where $k$ in the number of bits used to estimate the QBER. Then if the inequality
\begin{equation}\label{eq:pa}
	2^{-\frac15(l_{\rm ver}(1-h_\mathrm{b}(\delta+\nu))-r-t-l_{\rm sec}}\leq\varepsilon_{\rm pa}
\end{equation}
is satisfied, then the generation of a private key of length $l_{\rm sec}$ and security parameter $\varepsilon_{\rm pa}$ is possible. 
Here $r$ is the length of the syndrome in the error correction procedure and $t$ is the length of the verification hash.

If formula (\ref{eq:pa}) gives the positive answer on the question about the possibility of key generation with the desired parameters, 
then the final private key is computed as a hash function of the input bit string. 
The family of hash functions is required to be 2-universal \cite{Tomamichel}. 
The generalization to almost universal families of hash functions is also possible.

A hash function is chosen randomly from the Toeplitz universal family of hash functions \cite{Krawczyk}. 
A matrix $T$ of dimensionality $l_{\rm sec}\times l_{\rm ver}$ is a Toeplitz matrix if $T_{ij}=T_{i+1,j+1}=s_{j-i}$ for all $i=1,\ldots,l_{\rm sec}-1$ and $j=1,\ldots,l_{\rm ver}-1$. 
Thus, to generate a Toeplitz matrix, we need a random bit string $S=(s_{1-l_{\rm sec}},s_{1-l_{\rm sec}+1},\ldots,s_{l_{\rm ver}-1})$ of length $l_{\rm ver}+l_{\rm sec}-1$. 
The string $S$ is generated randomly by one side (say, Alice) and sent to another side (Bob) by public channel. Let us denote the corresponding Toeplitz matrix as $T_S$. 
Then the final private key is computed as $K_{\rm sec}=T_SK_{\rm ver}$ (with multiplication and addition modulo 2).

It is possible to use not a random, but a pseudo-random bit string, which uses a shorter random seed, to specify the Toeplitz matrix \cite{Krawczyk,Krawczyk2}. 
In this case, the family of hash functions is not universal, but almost universal, which is also acceptable for privacy amplification with some modifications in formula (\ref{eq:pa}). 
The security of privacy amplification is based on the Leftover hash lemma, which can be proved for almost universal hash functions as well \cite{TomRenner}. 
However, neither the rate of random number generator, nor the amount of publicly amount information are critical parameters of our setup. 
Thus, we adopt the standard family of Topelitz functions.

\section{Authentication}\label{sec:authentication}

The purpose of the authentication is to ensure that messages received by each side via public channel were sent by the other legitimate side (not by the adversary) 
and were not changed during the transmission.

Usual way to deal with this problem is hashing of the messages by hash functions dependent on a private key $K_{\rm aut}$ known only to legitimate parties. 
In general, the procedure is as follows: Alice sends to Bob a message with the hash tag. 
After receiving the message, Bob also computes the hash tag of the message and compare it to that received from Alice. 
If the hash tags coincide, Bob acknowledge the message as sent by Alice, otherwise he breaks the protocol. 
The hash function requires to assure that, whenever an eavesdropper does not know the private key, 
he cannot modify the message of send his own message and guess the correct hash tag of the message except for negligible probability 
(we require it not to exceed $\varepsilon_{\rm aut}=10^{-12}$).
 
For unconditional security, the hash function must be chosen from some universal family: 
almost strongly universal \cite{WegCar,Stinson}, almost xor-universal \cite{Krawczyk,Shoup}, or almost $\Delta$-universal family \cite{MMH}.

After consideration of many universal families of hash functions, we have decided in favour of the Toeplitz hashing (see Sec.~\ref{sec:pa} above) due to its computational simplicity. 
In the privacy amplification procedure, we exploit that it is a 2-universal family. 
Here we exploit its xor-universality \cite{Krawczyk}. 
Let the lengths of the authenticated messages and their hash tags be $l_M$ and $l_h$ respectively. 
The hash tag of the $i$th message $M_i$ is calculated as
\begin{equation}\label{eq:aut}
	h(M_i)=T_SM_i\oplus r_i,
\end{equation}
where $T_S$ is a $l_h\times l_M$ Toeplitz matrix generated by a string $S$ of length $l_h+l_M-1$, $r_i$ is a bit string of length $l_h$, and $\oplus$ is the bitwise xor. 
Both $S$ and $r_i$ are private and taken from the common private key $K_{\rm aut}$. 
Then, the probability that an eavesdropper will guess the hash tag of a modified message is not more than $2^{-l_h}$. The demand $\varepsilon_{\rm aut}=10^{-12}$ gives $l_h=40$.

As in privacy amplification, we do not adopt a pseudo-random generation of a string $S$ from a shorter string. 
This reduces the consumption rate of the private key. 
However, formula (\ref{eq:aut}) allows to generate the bit string $S$ only once, for the first message. 
For further messages, the private key is consumed in the rate $l_h$ bits per message (for strings $r_i$). 
Thus, large initial consumption of the private is not critical. 
The string $S$ may be used for many sessions of quantum key distribution.

Also this consideration suggests that it is advantageous to authenticate several messages at once: 
except for the first message, the consumption of the private key is $l_h$ bit per message and does not depend on the length of the message. 
However, the reduction of the number of authentications raises the risk of denial-of-service attacks from an eavesdropper: 
he is able to simulate messages from legitimate parties and force them to do calculations before his interference will be disclosed (in the nearest authentication stage). 
Thus, there is a trade-off between the private key consumption rate and the risk of denial-of-service attacks.

During the post-processing, the public classical channel is used  twice: 
(i) to send the syndrome in the error correction stage along with the verification hash tag and 
(ii) to send the decision about possibility of key generation and (if the answer is positive)
estimated level of QBER and bit string used to generate the Toeplitz matrix in the privacy amplification algorithm. 

Another  possibility is to use the hash function based on the known GOST cipher \cite{Schneier}, though it is not unconditionally secure.

\section{Workflow}\label{sec:workflow}

The workflow of the post-processing procedure is as follows. 
Sifted keys go through the error correction that is adjusted on the current value of QBER. 
After accumulation of necessary number of blocks they input to the parameter estimation (together with their versions before the error correction). 
If an estimated value of QBER given by (\ref{QBER}) is higher than the critical value needed for efficient privacy amplification, the parties receive warning message about possible of eavesdropping. 
Otherwise, verified blocks input privacy amplification and estimated QBER is used in next round of the error correction algorithm. 

The overall (in)security parameter of the quantum key distribution system is
\begin{equation}
\begin{split}
	\varepsilon_{\rm qkd}&=\varepsilon_{\rm ver}+\varepsilon_{\rm pa}+\varepsilon_{\rm aut}\\
	&=2\times10^{-11}+10^{-12}+10^{-12}<3\times10^{-11}.
\end{split}
\end{equation}
This parameter majorizes both the probability that the keys of Alice and Bob do not coincide and the probability of guessing the common key by Eve.
If this parameter exceeds a critical value, then the protocol is terminated.

After the privacy amplification procedure, a fraction of the key is used for authentication in the next rounds. 
In our setup, the fraction of generated private key consumed by authentication procedure does not exceed 15\%.

\medskip

\section{Conclusion}\label{sec:conclusion}

We have present post-processing procedure for industrial quantum key distribution systems.
The post-processing procedure consisting of error correction, parameter estimation, and privacy amplification has been described. 
Also authentication of classical communications over a public channel has been considered.

\section*{Acknowledgments}

We thank Andrey Fedchenko for useful comments. 
The support from Ministry of Education and Science of the Russian Federation in the framework of the Federal Program (Agreement 14.579.21.0104, ID RFMEFI57915X0104) is acknowledged. 
We thank the organizers of the 3rd International School and Conference Saint-Petersburg OPEN 2016 for kind hospitality.

\end{document}